\documentclass[aps,prb,amsmath,amssymb,superscriptaddress,notitlepage,twocolumn,reprint]{revtex4-1}
\usepackage{chngcntr}\usepackage{graphicx}
\usepackage{dcolumn}
\usepackage{stackengine}
\usepackage[section]{placeins}
\usepackage[dvipsnames*,svgnames]{xcolor}
\usepackage{verbatim}
\usepackage[position=bottom,caption=false,captionskip=3pt,font=normalsize,subrefformat=simple,labelformat=simple]{subfig}

\usepackage{listings}
\usepackage{mathtools}
\usepackage{amsmath}
\usepackage{chngcntr}
\usepackage{bm}
\usepackage{algorithm}
\usepackage{algpseudocode}
\usepackage{url}
\usepackage[super]{nth}
\usepackage{placeins}
\definecolor{GRgreen}{rgb}{0.2, 0.55, 0}
\newcommand{\AG}[1][]{\textcolor{red}}
\newcommand{\GR}[1][]{\textcolor{GRgreen}}
\newcommand{\AH}[1][]{\textcolor{blue}}
\usepackage{nicefrac}       
\usepackage{multirow}
\usepackage[version=3]{mhchem}
\usepackage{siunitx}
\usepackage{microtype}      
\usepackage{lipsum}

\usepackage[breaklinks]{hyperref}
\usepackage{tabularx}
\usepackage{soul}
\hypersetup{
colorlinks,
linkcolor={blue!100!black},
citecolor={blue},
urlcolor={blue!80!black}}

\newcommand\varpm{\mathbin{\vcenter{\hbox{%
  \oalign{\hfil$\scriptstyle+$\hfil\cr
          \noalign{\kern-.3ex}
          $\scriptscriptstyle({-})$\cr}%
}}}}
\newcommand\varmp{\mathbin{\vcenter{\hbox{%
  \oalign{$\scriptstyle({+})$\cr
          \noalign{\kern-.3ex}
          \hfil$\scriptscriptstyle-$\hfil\cr}%
}}}}

\usepackage{array}

\usepackage{booktabs}

\def\equationautorefname~#1\null{#1\null}
\usepackage[bottom]{footmisc} \makeatletter \def\p@section{} \def\p@subsubsection{} \makeatother
\begin{document}

\title{Mode- and Space- Resolved Thermal Transport of Alloy Nanostructures}

\author{S. Aria Hosseini}

\affiliation{Department of Chemistry, Massachusetts Institute of Technology, 77 Massachusetts Avenue,
Cambridge, Massachusetts 02139, USA}

\author{Sarah Khanniche}

\affiliation{Department of Chemical Engineering, Massachusetts Institute of Technology, 77 Massachusetts Avenue,
Cambridge, Massachusetts 02139, USA}

\author{G. Jeffrey Snyder}
\affiliation{Department of Materials Science and Engineering, Northwestern University, Evanston, IL, USA}

\author{Samuel Huberman}
\email{samuel.huberman@mcgill.ca}

\affiliation{Department of Chemical Engineering, University of McGill, 3610 University Street
Montreal, Quebec H3A 0C5}

\author{P. Alex Greaney}
\email{greaney@ucr.edu}

\affiliation{Department of Mechanical Engineering, University of California, Riverside, 900 University Avenue, Riverside, CA 92521, USA}

\author{Giuseppe Romano}
\email{romanog@mit.edu}

\affiliation{Institute for Soldier Nanotechnologies, Massachusetts Institute of Technology, 77 Massachusetts Avenue,
Cambridge, Massachusetts 02139, USA}

\begin{abstract}

Nanostructured semiconducting alloys obtain ultra-low thermal conductivity as a result of the scattering of phonons with a wide range of mean-free-paths (MFPs). In these materials, long-MFP phonons are scattered at the nanoscale boundaries whereas short-MFP high-frequency phonons are impeded by disordered point defects introduced by alloying.
While this trend has been validated by simplified analytical and numerical methods, an \emph{ab-initio} space-resolved approach remains elusive. To fill this gap, we calculate the thermal conductivity reduction in porous alloys by solving the mode-resolved Boltzmann transport equation for phonons using the finite-volume approach. We analyze different alloys, length-scales, concentrations, and temperatures, obtaining a very large reduction in the thermal conductivity over the entire configuration space. For example, a $\sim$97\% reduction is found for $\mathrm{Al_{0.8}In_{0.2}As}$ with 25\% porosity. Furthermore, we employ these simulations to validate our recently introduced ``Ballistic Correction Model'' (BCM), an approach that estimates the effective thermal conductivity using the characteristic MFP of the bulk alloy and the length-scale of the material. The BCM is then used to provide guiding principles in designing alloy-based nanostructures. Notably, it elucidates how porous alloys such as $\mathrm{Si_{x}Ge_{1-x}}$ obtain larger thermal conductivity reduction compared to porous $\mathrm{Si}$ or $\mathrm{Ge}$, while also explaining why we should not expect similar behavior in alloys such as $\mathrm{Al_{x}In_{1-x}As}$. By taking into account the synergy from scattering at different scales, we provide a route for the design of materials with ultra-low thermal conductivity.

\end{abstract}

\maketitle

\section{introduction}

Nanostructured semiconducting alloys may have ultra-low thermal conductivity: long-MFP phonons are scattered through nanostructuring and high-frequency phonons, which usually have short-to-medium MFPs, are scattered by mass disorder introduced by alloying. However, a first-principles mode- and space- resolved analysis of heat transport that can account for the interplay of the scattering processes in these systems remains elusive. To fill this gap, in this work we solve the space-dependent phonon Boltzmann transport equation (BTE) in porous materials (Sec.~\ref{BTE}). We consider a broad sweep of binary and ternary alloys of Group-IV and Group III-V and obtained a strong reduction in the effective thermal conductivity in all cases. For example, we compute a $\sim$97 \% degradation in $\mathrm{Al_{0.8}In_{0.2}As}$ with 25\% porosity compare to bulk non-alloy $\mathrm{AlAs}$. Mode-resolved analysis reveals phonon filtering over a wide portion of the MFP spectrum (Sec.~\ref{bandpass}). Furthermore, we validate the recently-developed ``Ballistic Correction Model'' (BCM)~\cite{hosseini2022universal} against BTE calculations (Sec.~\ref{BCM}), allowing the exploration of compounds beyond the ones studied in this work. The BCM enables us to provide guidelines for the holistic design of alloy-based nanostructures for thermoelectric applications. Finally, we study the reduction in thermal conductivity using BTE and BCM and explain why we should expect to obtain larger thermal conductivity reduction in porous alloys such as $\mathrm{Si_{x}Ge_{1-x}}$ compared to non-alloys counterparts (e.g., $\mathrm{Si}$ or $\mathrm{Ge}$), while we should not expect similar behavior in alloys such as $\mathrm{Al_{x}In_{1-x}As}$.

Nanoengineered membranes including thin films,~\cite{cheaito2012experimental,braun2016size} 
nanowires,~\cite{li2012thermal} nanomeshes,~\cite{feng2016ultra,perez2016ultra} nanocomposites,~\cite{miura2015crystalline,liao2015nanocomposites} and nanoporous structures \cite{doi:10.1021/acsaem.0c02640,shi2018polycrystalline,PhysRevB.100.035409,PhysRevB.102.205405} feature extremely low thermal conductivity, sometimes beyond the amorphous limit.~\cite{lim2016simultaneous} This is largely due to the scattering of phonons with large MFPs (generally with small frequencies), i.e., those that are comparable with the characteristic length of the material.
Low-thermal conductivity can also be obtained with point defects~\cite{gurunathan2020analytical, al2016band}, such as those induced by alloying,~\cite{gurunathan2020alloy, lim2016simultaneous} and vacancies~\cite{PhysRevB.102.205405}. Point-defect scattering strongly suppresses medium- to high- frequency phonons, mostly with shorter MFPs while leaving low-frequency phonons (generally with longer MFPs) unimpeded. 
Thus, the combination of alloy and porosity scattering leads to extremely short-MFP across a wide range of phonon frequencies. Ultra-low thermal conductivity has been, in fact, observed in several experiments, including nanoporous $\mathrm{Si_x Ge_{1-x}}$,~\cite{perez2016ultra} and porous $\mathrm{Sn Se_x S_{1-x}}$ Nano-sheets~\cite{doi:10.1021/acs.chemmater.7b00423}.

\begin{figure*}[t]

    \centering
    \subfloat[\label{fig:1.1}]{\includegraphics[width=0.48\textwidth]{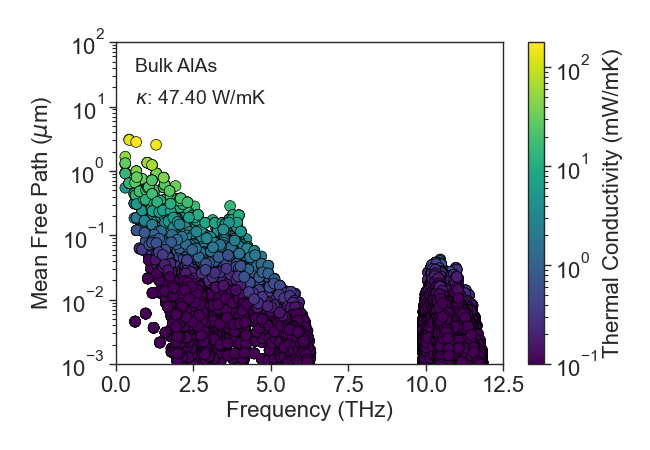}}
    \centering
    \subfloat[\label{fig:1.2}]{\includegraphics[width=0.48\textwidth]{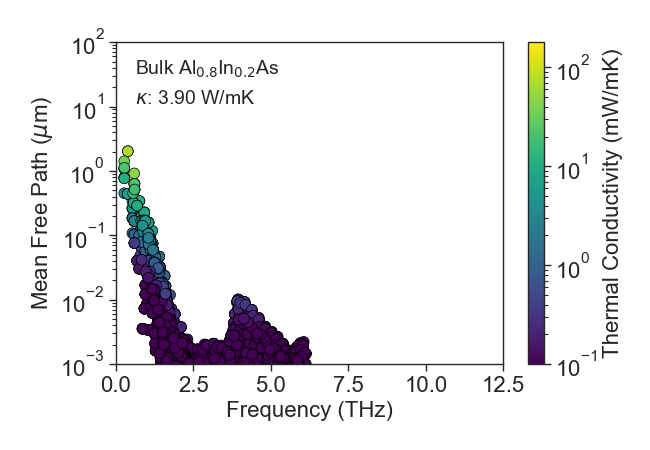}}
    
    \centering
    \subfloat[\label{fig:1.3}]{\includegraphics[width=0.48\textwidth]{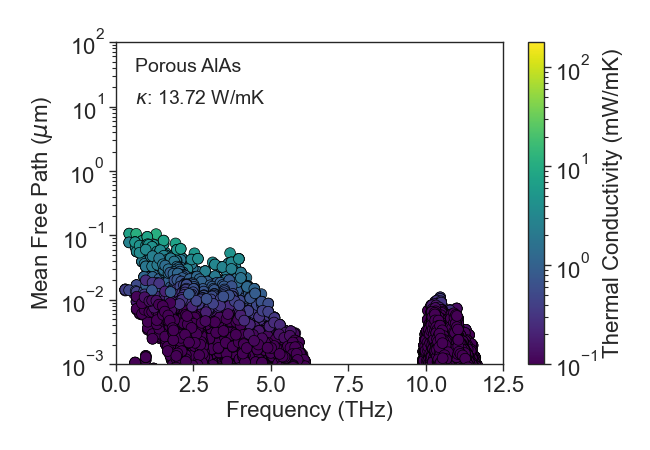}}
    \centering
    \subfloat[\label{fig:1.4}]{\includegraphics[width=0.48\textwidth]{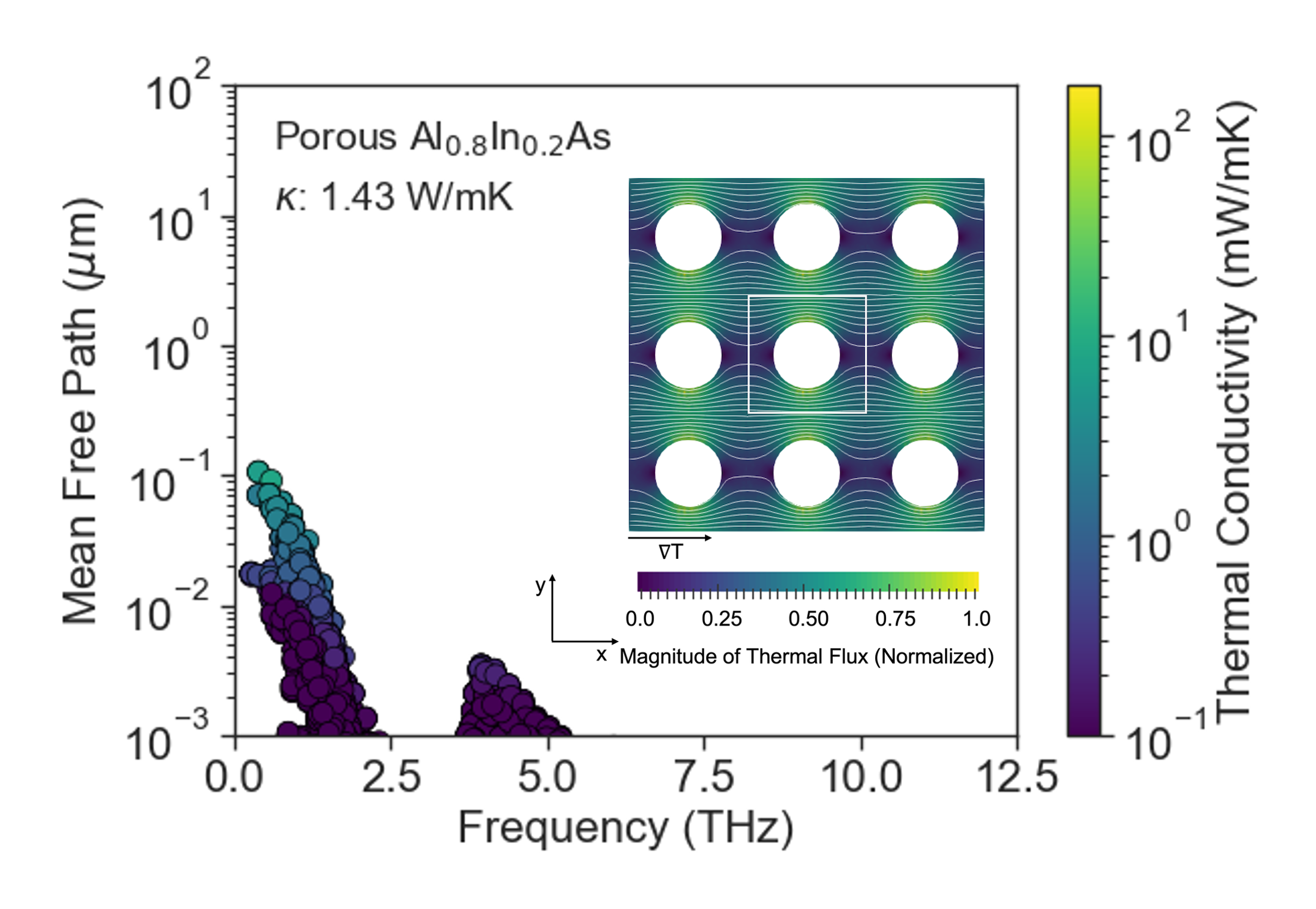}}
    
    \caption{Phonon suppression via nanostructuring. Panels (a) and (b) show mode-resolved mean free path of bulk AlAs and Al\textsubscript{0.8}In\textsubscript{0.2}As, respectively. High-frequency phonons are completely filtered out in the alloy. Plot (c) shows the same material as in (a) but with nanoscale cylindrical porosity. Porosity is fixed at $\mathrm{\phi=0.25}$ and pores are aligned and 100 nm apart (center to center distance). Long MFP phonons are strongly suppressed at the surfaces of the pores. Pane (d) shows the synergy effect of alloy and pore scatterings on lattice thermal conductivity. Both low-frequency high-MFP phonons and high-frequency low-MFP phonons are suppressed leading to ultra-low thermal conductivity. The thermal conductivity is printed top right of each figure. The temperature is fixed at 500 K. (Inset) Magnitude of the thermal flux in porous Al\textsubscript{0.8}In\textsubscript{0.2}As with aligned cylindrical pores. The porosity is fixed at $\phi=0.25$, and the spacing between pores is $L=50\ \mathrm{nm}$. The white box shows the borders of the unit volume. The heat flux is higher in the constriction between the pores.} 
    \label{fig:1}
    
\end{figure*}

Several aspects of alloy-based nanostructures have been studied computationally. For example, MC simulations, with phenomenological input data, have shown that nanostructuring can be more effective in Si$_{0.5}$Ge$_{0.5}$ alloys than in their end-compounds counterpart~\cite{bera2010marked}. In another study, Yang et al.~\cite{yang2017thermal} investigated thermal transport in nano-crystalline SiGe, with bulk data computed by first principles. In this work, we combine these two approaches and investigate space- and mode- resolved thermal transport in alloy-based nanostructures from first principles. To this end, we employ the recently developed anisotropic MFP-BTE~\cite{romano2021}, which takes into account the whole MFP spectrum and therefore captures phonons interactions in the non-diffusive regime.~\cite{harter2019prediction, harter2020predicting}

To accelerate the screening of these systems, we assess the accuracy of the recently introduced ``Ballistic Correction Model'' (BCM)~\cite{hosseini2022universal}; the BCM is an effective medium theory that provides a closed-form model for estimating the thermal conductivity by taking into account the whole bulk MFP distribution as well as the MFP-dependent phonon suppression induced by a given geometry. In practice, these two quantities are approximated by a logistic function, with parameters being the characteristic MFP and length scale. In this work, we validate the BCM against the BTE for alloys in groups IV and III-V and for different length scales and tabulate the corresponding parameters. Lastly, the BCM is used to determine when alloying yields a relaxation in the nanostructure resolution to achieve a given thermal conductivity reduction.

\begin{figure*}[t]
        
    \centering
    \subfloat[\label{fig:3.1}]{\includegraphics[width=0.48\textwidth]{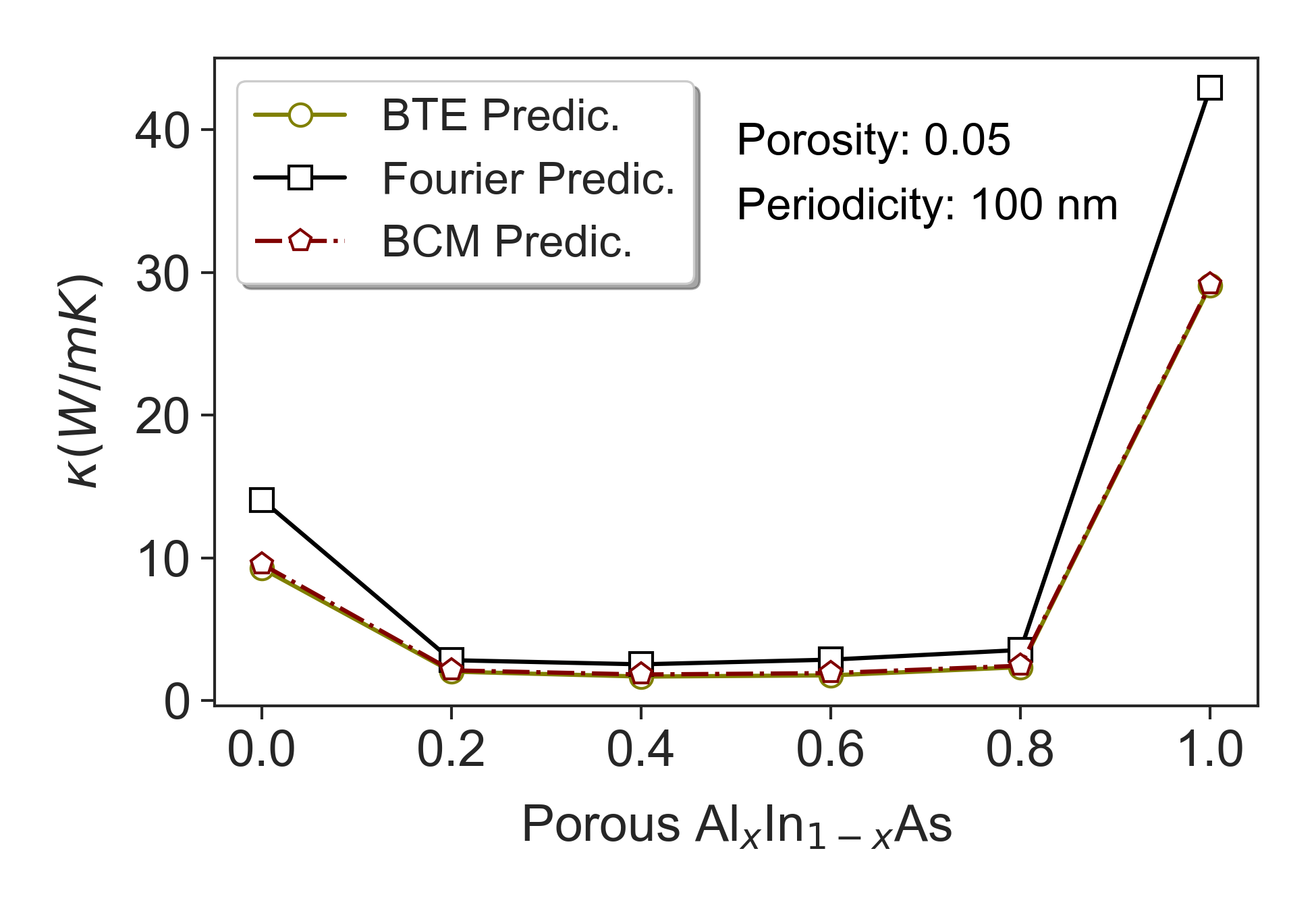}}
    \centering
    \subfloat[\label{fig:3.2}]{\includegraphics[width=0.48\textwidth]{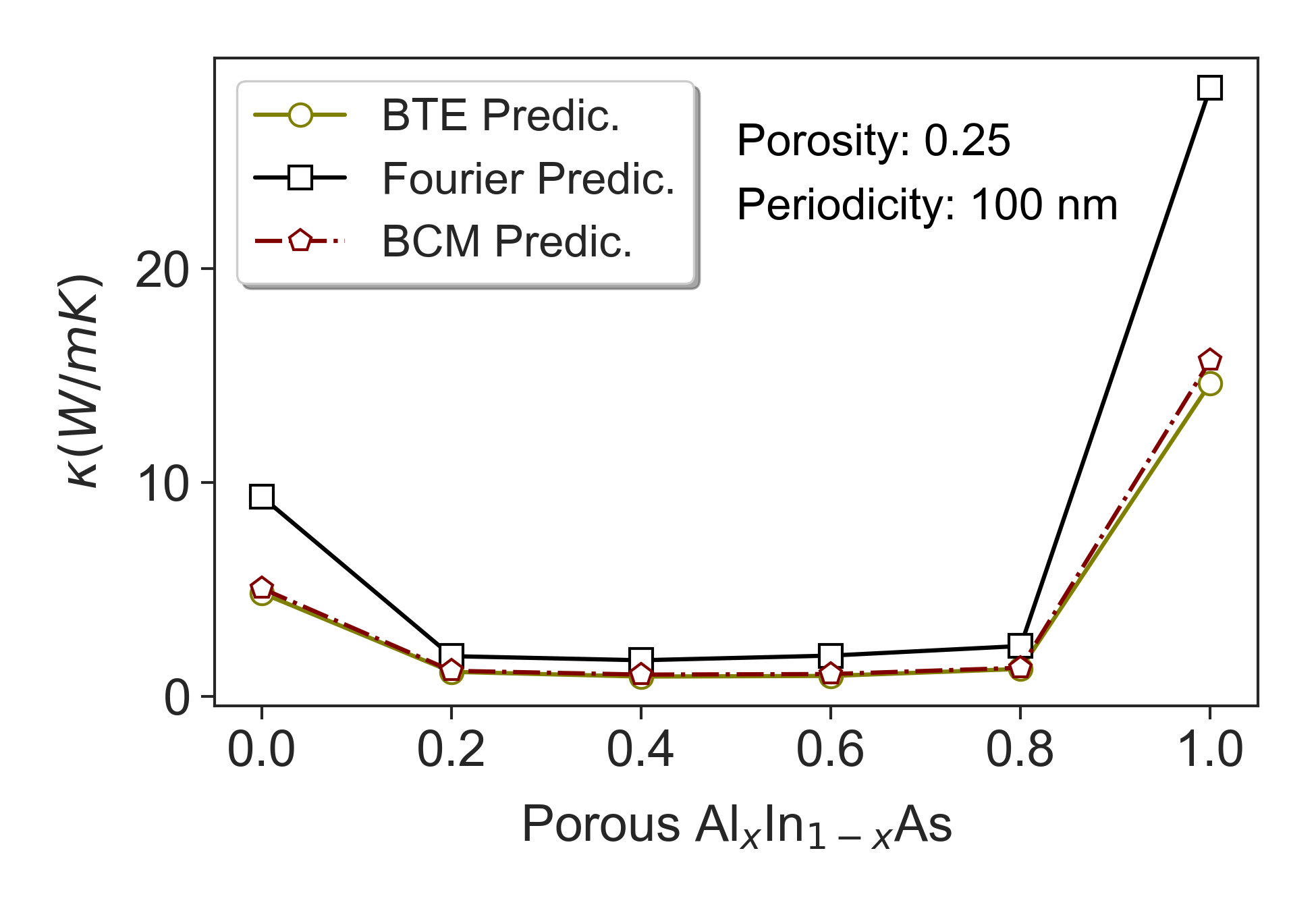}}
    
    \centering
    \subfloat[\label{fig:3.3}]{\includegraphics[width=0.48\textwidth]{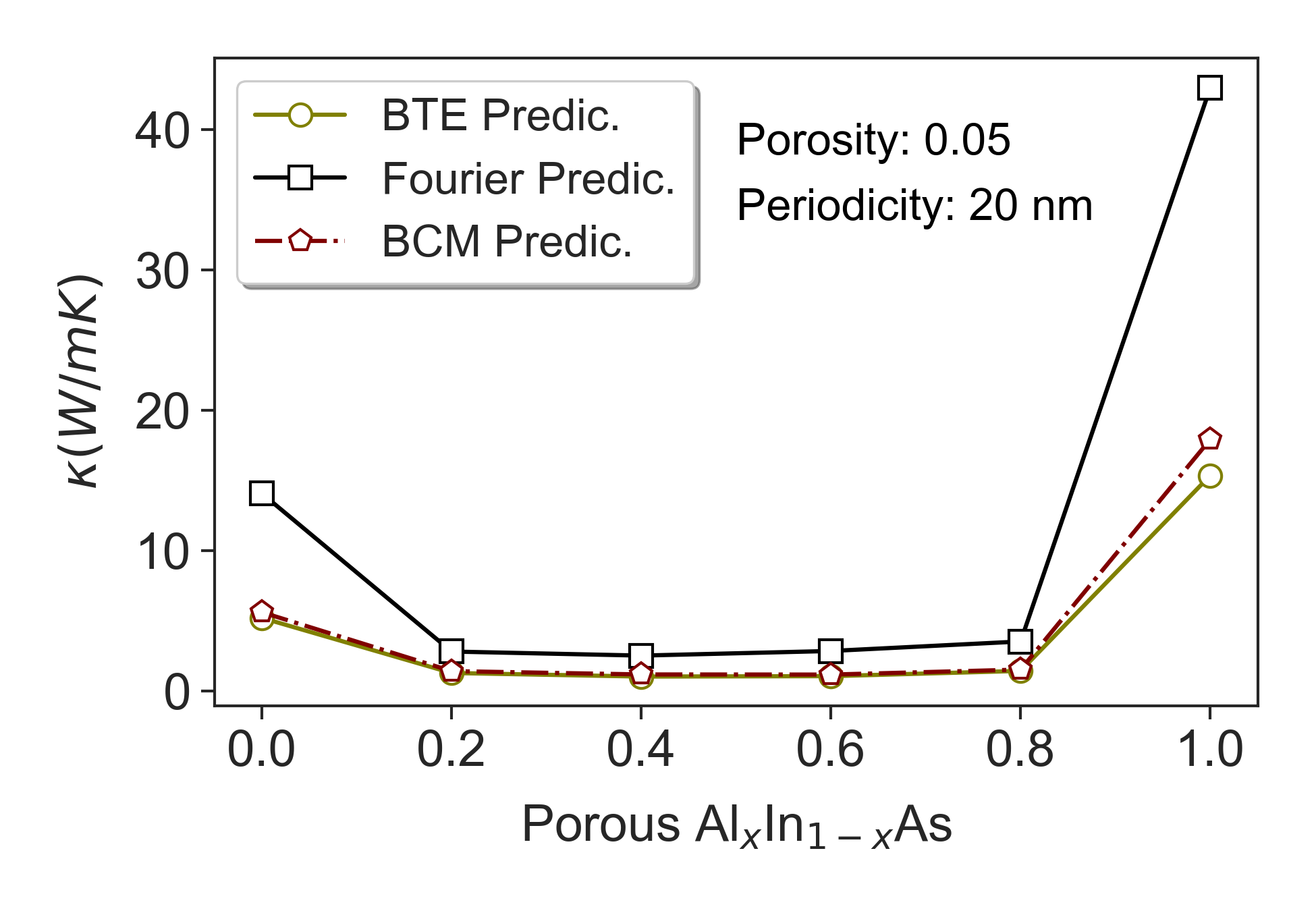}}
    \centering
    \subfloat[\label{fig:3.4}]{\includegraphics[width=0.48\textwidth]{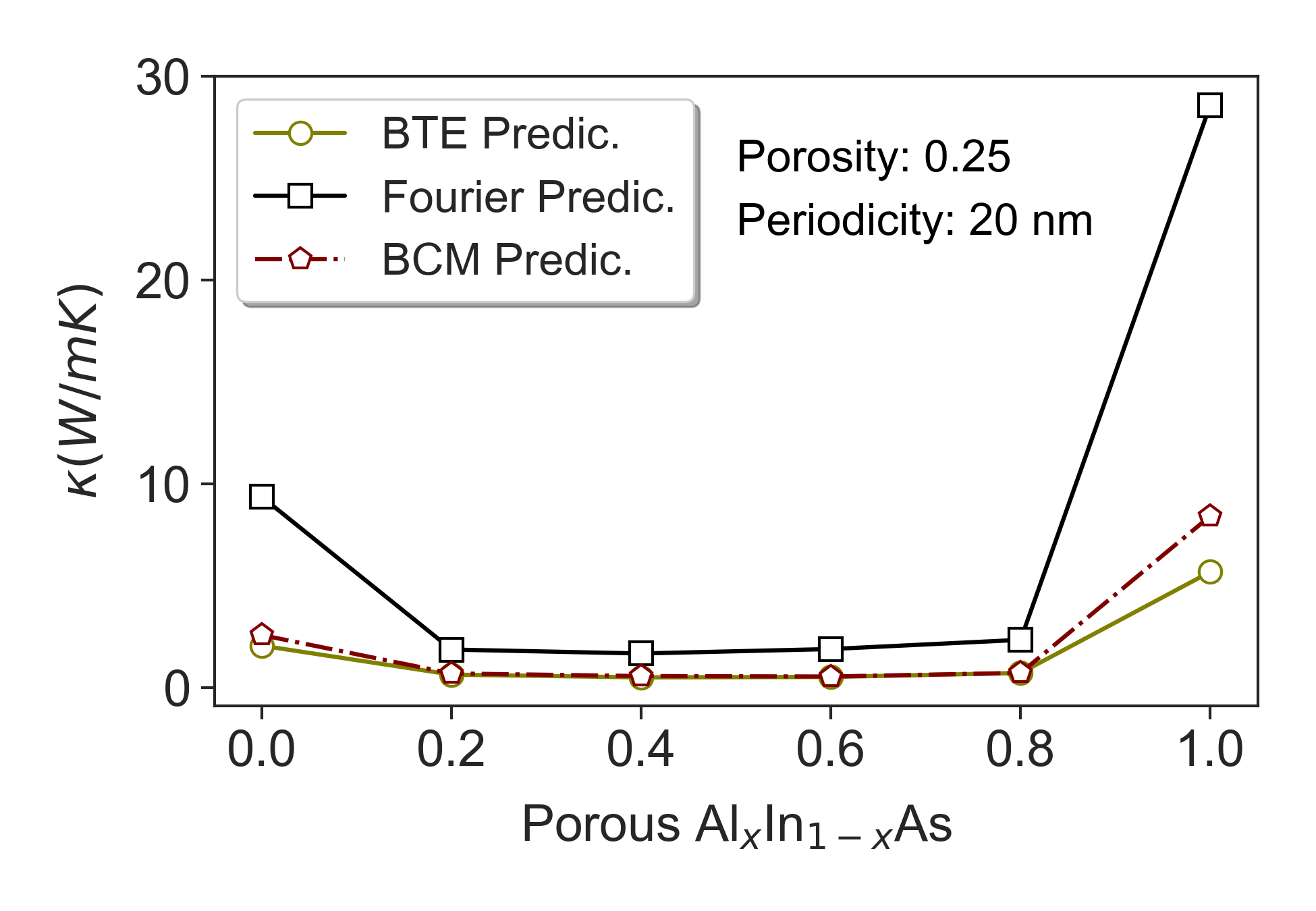}}
    
    \caption{The thermal conductivity predictions by BTE, BCM, and Fourier's law in AlInAs at the fixed temperature of 500 K. The porosity is fixed at 0.25 in (a) \& (c) and at 0.05 in (b) \& (d). The periodicity is 100 nm in (a) \& (b) and 20 nm in (c) \& (d). The green lines with open circles show the BTE prediction, the black lines with open squares markers show Fourier's predictions, and BCM estimations are plotted in red dot-dash lines with pentagon markers. The bulk characteristic MFPs in AlInAs alloys are about 100 nm. This is comparable with the periodicity in (a) and (b). In this regime, all scattering processes including phonon-phonon, phonon mass-disorder, and phonon-boundary are important. In panels (c) and (d) the length scale is small enough that only phonon mass-disorder and phonon-boundary scatterings contribute to the bulk thermal conductivity. In all cases, the transport is non-diffusive. The BCM prediction is in excellent agreement with BTE for both length scales.}
    
    \label{fig:3}
\end{figure*}
\section{The BTE-based approach}\label{BTE}


Space-dependent simulations of nanostructured alloys employed in this work are based on the Boltzmann transport equation (BTE) approach.~\cite{ziman1972principles} Specifically, we employ the recently-developed anisotropic MFP-BTE,~\cite{romano2021} implemented in OpenBTE,
\begin{equation}\label{bte}
\sum_{\mu'}\left[\gamma \frac{ C_{\mu'}}{\tau_{\mu'}} -\delta_{\mu\mu'}\left(1+ \tau_{\mu'}\mathbf{v}_{\mu'}\cdot\nabla\right)\right]\Delta T_{\mu'}=0,
\end{equation}
where $\gamma=\left(\sum_k C_k/\tau_k\right)^{-1}$, $C$ is the mode-resolved heat capacity and $\mathbf{v}_\mu$ is the phonon group velocity. Equation~\eqref{bte} must be solved for each phonon mode. In practice, we employ interpolation in angular and MFP-space.~\cite{romano2021} The simulation domain is a two-dimensional square and lies on the $xy$-plane with one circular pore located at its center- see inset in Fig.~\ref{fig:1}; the model is translational invariant along the $z$-axis thus the actual system is infinite along the out-of-plane direction. Periodic boundary conditions are applied along both axes, and a temperature difference of 1 K is applied at two opposite sides perpendicular to the $x$-axis. The pore/dielectric interface is modeled as a diffusely scattering adiabatic boundary. This means that the total energy flux incident on an interface is re-emitted back into the simulation domain in all directions distributed over all phonon frequencies in proportion to their equilibrium occupancy. Throughout the text, the geometry is parametrized by the porosity $\phi$, i.e., the volume fraction between the alloy and the dielectric, and the periodicity $L$, which is simply the size of the unit cell. For all the configurations used in this work, we ensure that the smallest pore spacing is larger than 10 nm, so that confinement effects are not significant, and that the phonon dispersion of the base material could still be reasonably approximated by those of the bulk crystal. 

Once Eq.~\eqref{bte} is solved, we compute the mode-resolved effective thermal conductivity using Fourier's law, i.e. 
\begin{equation}
    \kappa_\mu^{\mathrm{eff}} = -\frac{1}{\Delta T}\int_0^L \mathbf{J}_\mu\cdot \mathbf{\hat{n}} dy,
\end{equation}
where $\mathbf{J}_\mu = \mathcal{V}^{-1} C_\mu \mathbf{v}_\mu\Delta T_\mu $, with $\mathcal{V}$ being a normalization volume. The terms $C_\mu,\tau_\mu$ and $\mathbf{v}_\mu$ are computed by density functional theory. The second- and third- order interatomic force constants of bulk alloys, computed using virtual crystal approximation, are obtained from the AlmaBTE materials database.~\cite{carrete2017almabte} The phonon dispersion is computed from the second-order force constants on a $30\times30\times30$ point Brillouin zone mesh. The scattering rates for three-phonon interactions are computed from the third-order force constants. The rate of elastic phonon scattering by disordered atoms was modeled by treating them as random mass perturbations with the scattering rate given by Tamura's formula.~\cite{tamura1983isotope} This method does not account for correlations, local relaxations, or changes in electronic structure due to alloying and interatomic force constant disorder, yet gives a reasonable prediction for bulk thermal properties in most cases.~\cite{arrigoni2018first}

\section{PHONON SUPPRESSION THROUGH NANOSTRUCTURING} \label{bandpass}

\begin{figure}[t]
\centering
\includegraphics[width=0.48\textwidth]{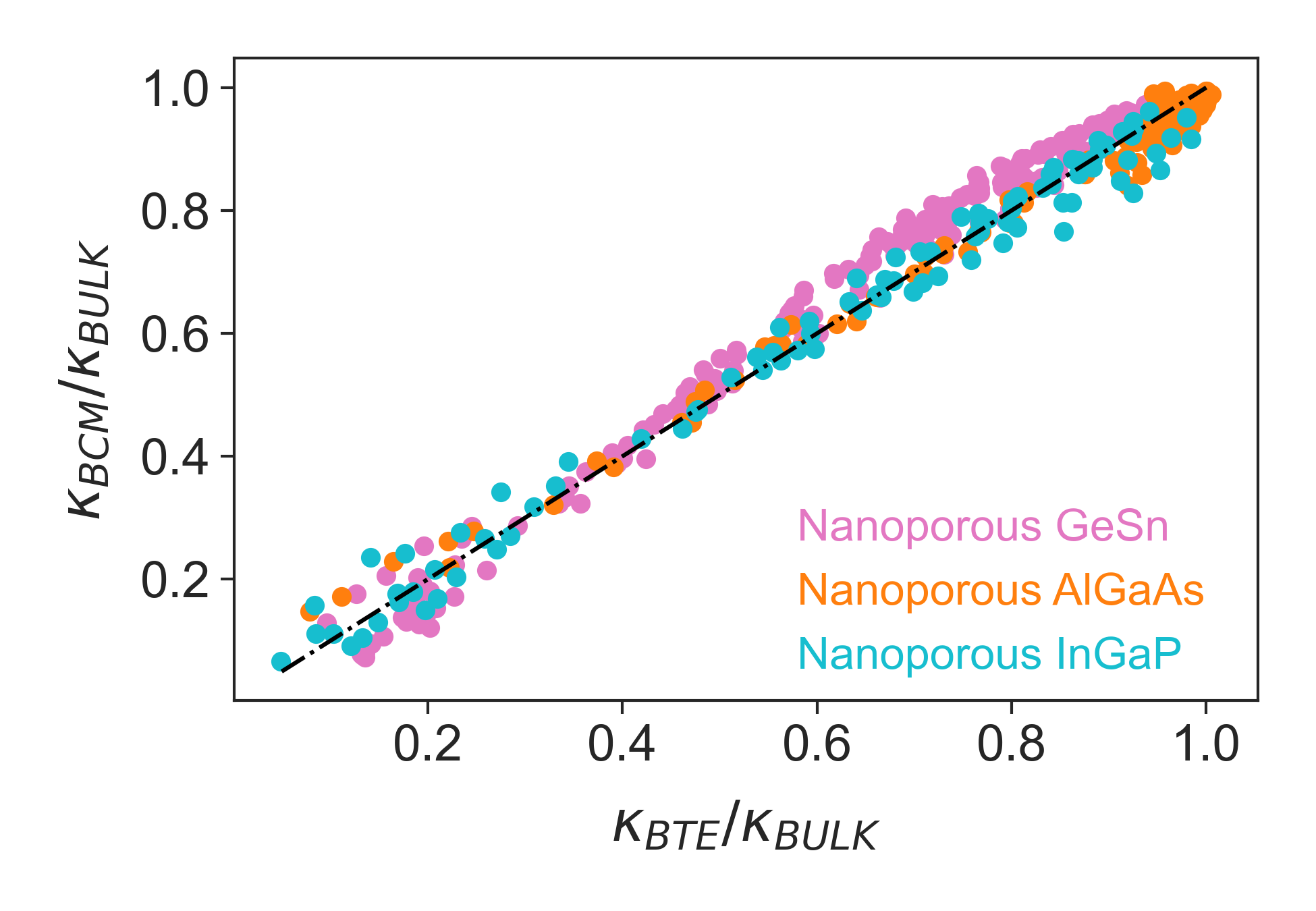}
\caption{Normalized thermal conductivity predicted by the reduced-order BCM \emph{vs.} BTE simulations for a selection of three base materials (Ge\textsubscript{x}Sn\textsubscript{1-x}, Al\textsubscript{x}Ga\textsubscript{1-x}As and In\textsubscript{x}Ga\textsubscript{1-x}P) with nanoscale pores with two different porosities of 0.05 and 0.25 and a variety of spacings. The periodicity varies from 10 nm up to 30 microns and the temperature varies from 300--800 K.} 
    \label{fig:5}
\end{figure}

The lattice thermal conductivity of alloys is generally determined by a combination of phonon-phonon Umklapp and mass-disorder scatterings.~\cite{kim2015strategies} The three-phonon scattering processes strongly rely on stiffness matrix and anharmonic bonding, while the mass-disorder relies on the difference in the atomic mass.~\cite{li2014shengbte} 
The mass mismatch strongly scatters high-frequency phonons~\cite{xie2013beneficial}, which generally have low MFPs,
while nanoscale porosity limits the transport of phonons with longer MFPs. Thus, alloys with nanoscale porosity imprint ultra-low thermal conductivity and suppresses phonons across the frequency spectrum.
The central strategy behind the phonon's filtering is to selectively scatter phonons within specific frequency ranges. Figure~\ref{fig:1} shows the phonon suppression due to the additive point defect alloys and nanoscale porosity in AlAs from first-principles calculations. The mode-resolved MFP of bulk AlAs and Al\textsubscript{0.8}In\textsubscript{0.2}As are shown in Figs.~\ref{fig:1.1} and~\ref{fig:1.2}, respectively. High-frequency optical phonons are completely filtered out and the acoustic phonons (except for the ones with a very low frequency) are strongly suppressed due to mass-disorder scattering. The reduction in thermal conductivity exceeds 90\%. Figure~\ref{fig:1.3} shows bulk AlAs with $\phi$ = 0.25 and L = 100 nm. More than 70\% reduction in lattice thermal conductivity through filtering long-MFP phonons is observed. 
As mentioned above, pores have a weaker impact on phonons with medium to high frequency because the intrinsic MFPs are generally smaller.
Figure~\ref{fig:1.3} shows the cumulative effect of alloying and nanoscale porosity. Phonons across the whole frequency spectrum are strongly suppressed leading to $\sim$ 97\% reduction in thermal conductivity. Such an extreme reduction in thermal conductivity is due to the interplay between phonon-boundary and phonon mass mismatch scatterings~\cite{toberer2011phonon}.   Figure~\ref{fig:3} illustrates the effective thermal conductivity of Al\textsubscript{x}In\textsubscript{1-x}As at 500 K computed by both the BTE and the standard heat conduction equation, i.e. $\nabla^2 \Delta T = 0$.  
Two sets of porosities are considered, $\mathrm{\phi = 0.05}$ and $\mathrm{\phi = 0.25}$. In Figs.~\ref{fig:3.1} and~\ref{fig:3.2}, the periodicity is fixed at 100 nm. This is comparable with characteristic MFP in AlInAs alloys; thus, all scattering processes contribute to the bulk thermal conductivity. In Figs.~\ref{fig:3.3} and~\ref{fig:3.4}, the periodicity is fixed at 20 nm. For this length-scale, mass-disorder and phonon-boundary scattering dominate over three-phonon scattering. In all cases, the transport is non-diffusive.

\section{Ballistic Correction Model}\label{BCM}

\begin{figure*}[t]

    \centering
    \subfloat[\label{fig:7-1}]{\includegraphics[width=0.48\textwidth]{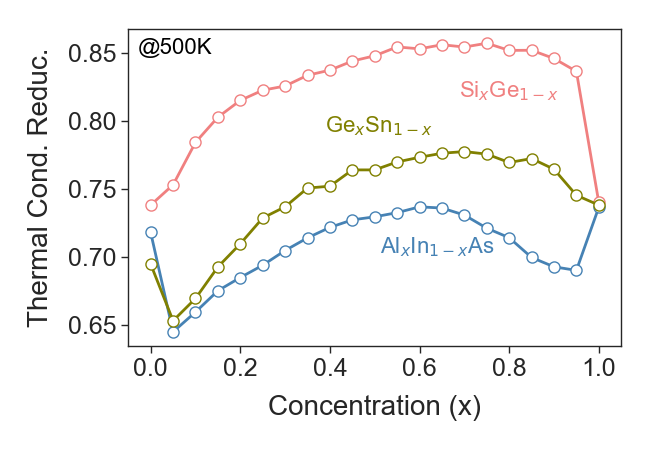}}
    \centering
    \subfloat[\label{fig:7-2}]{\includegraphics[width=0.48\textwidth]{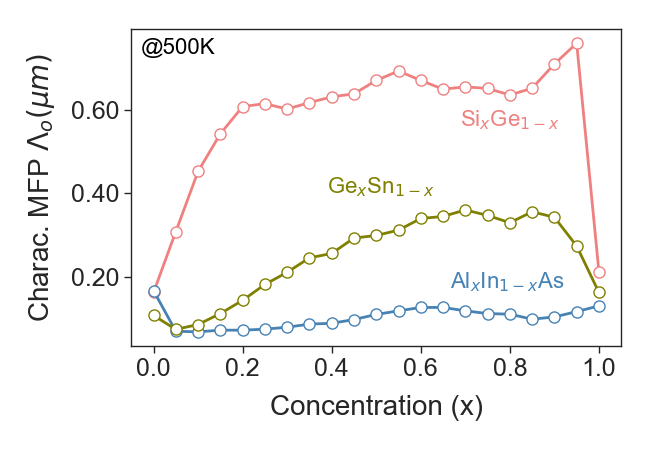}}
    
    \caption{The thermal conductivity reduction predicted by BTE simulations (a) and the characteristic MFP predicted by BCM (b) for a selection of three base materials (Ge\textsubscript{x}Sn\textsubscript{1-x}, Al\textsubscript{x}In\textsubscript{1-x}As and Ge\textsubscript{x}Sn\textsubscript{1-x}) with 0.25 porosities and pore-pore spacing of 100 nm. The temperature is fixed at 500 K. The $\Lambda_o$ of SiGe alloys are higher than the end-compounds while $\Lambda_o$ of AlInAs alloys are shorter than the end-compounds.}
    
    \label{fig:7}
\end{figure*}

We begin this section by briefly describing the ``Ballistic Correction Model" (BCM), introduced in our previous work.~\cite{hosseini2022universal} Given a geometry (which here is defined as a scale-independent entity), and the underlying bulk material, the  BCM estimates the effective thermal conductivity as
\begin{equation}\label{eq:keff}
\kappa_{\mathrm{eff}} = \kappa_{\mathrm{bulk}}S(0) \Xi(\mathrm{Kn}),
\end{equation}
where $\kappa_{\mathrm{bulk}}$ is the bulk thermal conductivity, $S(0)$ is the macroscopic suppression, and the ``Ballistic Correction Term" $\Xi(\mathrm{Kn})$ is given by
\begin{equation}\label{eq:Xi}
    \Xi (\mathrm{Kn}) = \left[\frac{1 + \mathrm{Kn} \left( \ln(\mathrm{Kn}) - 1  \right)}{\left ( \mathrm{Kn} -1 \right)^2} \right];
\end{equation}
in Eq.~\eqref{eq:Xi} $\mathrm{Kn}=\nicefrac{\Lambda_0}{\gamma L}$ is the Knudsen number, where $\Lambda_0$, $L$ is the periodicity and $\gamma$ is the normalized (by the periodicity) feature size of the nanostructure. The latter term as well as $S(0)$ are obtained by fitting the suppression function of a given geometry, obtained by material-free BTE calculations,~\cite{romano2019diffusive} against the logistic function $S(\xi)\approx S(0)\left[1+\xi/\gamma \right]^{-1}$. On the other side, $\Lambda_0$ is the characteristic MFP of the bulk material, obtained by fitting the normalized cumulative thermal conductivity against a logistic function $\alpha(\Lambda) = \left[1+\Lambda_0/\Lambda \right]^{-1} $. The steps for evaluating Eq.~\eqref{eq:keff} are as follows:~\cite{hosseini2022universal} (1) retrieve $\Lambda_0$ and $\kappa_{\mathrm{bulk}}$ for a given material and temperature, either from tabulated values,~\cite{Hosseini2021} or from new first-principles calculations, (2) retrieve $S(0)$ and $\gamma$ for a given geometry, either from a look-up table or from new BTE simulations, (3) set the periodicity of your system, $L$ (4) compute the Knudsen number and, finally, (5) evaluate Eqs.~\eqref{eq:Xi}-\eqref{eq:keff}.

The BCM prediction of thermal conductivity for AlInAs-based alloys is shown in Fig.~\ref{fig:3}.  The characteristic MFPs of AlAs, InAs, and Al\textsubscript{0.8}In\textsubscript{0.2}As are 0.13 micron, 0.17 micron, and 0.11 micron, respectively.~\cite{hosseini2022universal} The diffusive suppression factor $S(0)$ for 5\% and 25\% porosities are 0.90 \& 0.59, and the fractional characteristic lengths ($\gamma$) are 3.92 \& 1.82, respectively. As mentioned in the previous section, in Fig.~\ref{fig:3.1} and Fig.~\ref{fig:3.2} $L$ is comparable with $\Lambda_o$ of AlInAs alloys and in Fig.~\ref{fig:3.3} and Fig.~\ref{fig:3.4} $L$ is considerably smaller than the $\Lambda_o$ of AlInAs alloys. Yet, the model yields an excellent agreement for both length-scales. A similar agreement is found for a wider set of materials (GeSn, AlGaAs, and InGaP), with pore-pore spacing ranging from 10 nm up to 30 microns and temperature from 300-800 K (Fig.~\ref{fig:5}).  The cumulative thermal conductivity of the materials considered here as well as their least-squared logistic regression fit are available in the GitHub repository.~\cite{Hosseini2021}


BTE simulations unveil about 85\% reduction in thermal conductivity of $\mathrm{Si_{0.8}Ge_{0.2}}$ for 0.25 porosity and pore-pore spacing of 100 nm. On the other side, calculations for the same configuration, using Si as the base material, reveal a less extreme reduction of 74\%. This is in agreement with Ref.~\cite{bera2010marked}; However, this behavior is not general, e.g., the thermal conductivity reduction of $\mathrm{Al_{0.9}In_{0.1}As}$ and $\mathrm{AlAs}$ for the similar 0.25 porosity and 100 nm pore spacing, are 69\% and 73\%, respectively. Similarly, 67\% and 69\% reduction is observed in $\mathrm{Ge_{0.1}Sn_{0.9}}$ and $\mathrm{Sn}$, respectively [Fig.~\ref{fig:7-1}]. The reason for this opposite trend lies in the width of the MFP spectrum of the bulk end-compounds. For materials with large intrinsic MFPs, such as Si or Ge, the addition of alloy scattering reduces the small and medium MFPs, leaving, large MFP unaltered. Therefore, nanostructuring remains very effective and may lead, such as in the SiGe case, to a stronger thermal conductivity reduction with respect to the end-compounds counterparts, due to the predominance of large-MFP phonons. In terms of BCM, this scenario is captured by an increase of characteristic MFP $\Lambda_0$ with alloy concentrations, as shown in figure~\ref{fig:7-2}. We remark that $\Lambda_o$ is the median MFP of the thermal conductivity distribution and roughly the feature size in porous structures where the phonon-pore scattering takes precedence over the anharmonic/ mass-disorder scattering. Therefore, we expect a larger reduction in thermal conductivity for materials with higher $\Lambda_o$. On the other side, in materials with low MFPs, such as AlAs/ InAs, alloying shifts the entire spectrum toward smaller MFPs, making nanostructuring less effective. In this case $\Lambda_o$ decreases with the alloy concentration. In the case of GeSn, we observe both trends, as shown in Fig.~\ref{fig:7-1}. In fact, for a low concentration of Sn, the MFP spectrum is approximately as large as Ge's, thus we expect a larger reduction in thermal conductivity; however, for the high concentration of Sn, the intrinsic MFP spectrum becomes narrower (i.e., similar to the one of Sn), and the thermal conductivity reduction is smaller. The normalized thermal conductivity of SiGe, AlInAs, and GeSn are plotted in the SI.


\section{Conclusion}

To summarize, we solve the ab-initio Boltzmann transport equation to estimate the thermal transport in alloys with nanoscale porosity; our mode- and space- resolved analysis reveals how a synergetic effect of alloy and boundary scatterings leads to strong phonon filtering and thus ultra-low lattice thermal conductivity. Furthermore, we assess the accuracy of our recently-developed reduced-order approach, termed the ``Ballistic Correction Model'' (BCM),~\cite{hosseini2022universal} in estimating thermal transport in these systems, obtaining excellent agreement across a wide range of length scales and materials. Lastly, we provide guidelines for engineering these systems, using the BCM model and tabulated data. Providing a mode-resolved analysis of thermal transport in nanostructures alloys, our work may open up opportunities for a concurrent mode- and space- resolved engineering of low-thermal conductivity systems. 

\bibliographystyle{apsrev4-2}
\bibliography{references} 

\begin{thebibliography}{32}%
\makeatletter
\providecommand \@ifxundefined [1]{%
 \@ifx{#1\undefined}
}%
\providecommand \@ifnum [1]{%
 \ifnum #1\expandafter \@firstoftwo
 \else \expandafter \@secondoftwo
 \fi
}%
\providecommand \@ifx [1]{%
 \ifx #1\expandafter \@firstoftwo
 \else \expandafter \@secondoftwo
 \fi
}%
\providecommand \natexlab [1]{#1}%
\providecommand \enquote  [1]{``#1''}%
\providecommand \bibnamefont  [1]{#1}%
\providecommand \bibfnamefont [1]{#1}%
\providecommand \citenamefont [1]{#1}%
\providecommand \href@noop [0]{\@secondoftwo}%
\providecommand \href [0]{\begingroup \@sanitize@url \@href}%
\providecommand \@href[1]{\@@startlink{#1}\@@href}%
\providecommand \@@href[1]{\endgroup#1\@@endlink}%
\providecommand \@sanitize@url [0]{\catcode `\\12\catcode `\$12\catcode
  `\&12\catcode `\#12\catcode `\^12\catcode `\_12\catcode `\%12\relax}%
\providecommand \@@startlink[1]{}%
\providecommand \@@endlink[0]{}%
\providecommand \url  [0]{\begingroup\@sanitize@url \@url }%
\providecommand \@url [1]{\endgroup\@href {#1}{\urlprefix }}%
\providecommand \urlprefix  [0]{URL }%
\providecommand \Eprint [0]{\href }%
\providecommand \doibase [0]{https://doi.org/}%
\providecommand \selectlanguage [0]{\@gobble}%
\providecommand \bibinfo  [0]{\@secondoftwo}%
\providecommand \bibfield  [0]{\@secondoftwo}%
\providecommand \translation [1]{[#1]}%
\providecommand \BibitemOpen [0]{}%
\providecommand \bibitemStop [0]{}%
\providecommand \bibitemNoStop [0]{.\EOS\space}%
\providecommand \EOS [0]{\spacefactor3000\relax}%
\providecommand \BibitemShut  [1]{\csname bibitem#1\endcsname}%
\let\auto@bib@innerbib\@empty
\bibitem [{\citenamefont {Hosseini}\ \emph {et~al.}(2022)\citenamefont
  {Hosseini}, \citenamefont {Khanniche}, \citenamefont {Greaney},\ and\
  \citenamefont {Romano}}]{hosseini2022universal}%
  \BibitemOpen
  \bibfield  {author} {\bibinfo {author} {\bibfnamefont {S.~A.}\ \bibnamefont
  {Hosseini}}, \bibinfo {author} {\bibfnamefont {S.}~\bibnamefont {Khanniche}},
  \bibinfo {author} {\bibfnamefont {P.~A.}\ \bibnamefont {Greaney}},\ and\
  \bibinfo {author} {\bibfnamefont {G.}~\bibnamefont {Romano}},\ }\href
  {https://doi.org/https://doi.org/10.1016/j.ijheatmasstransfer.2021.122040}
  {\bibfield  {journal} {\bibinfo  {journal} {Int. J. Heat Mass Transf.}\
  }\textbf {\bibinfo {volume} {183}},\ \bibinfo {pages} {122040} (\bibinfo
  {year} {2022})}\BibitemShut {NoStop}%
\bibitem [{\citenamefont {Cheaito}\ \emph {et~al.}(2012)\citenamefont
  {Cheaito}, \citenamefont {Duda}, \citenamefont {Beechem}, \citenamefont
  {Hattar}, \citenamefont {Ihlefeld}, \citenamefont {Medlin}, \citenamefont
  {Rodriguez}, \citenamefont {Campion}, \citenamefont {Piekos},\ and\
  \citenamefont {Hopkins}}]{cheaito2012experimental}%
  \BibitemOpen
  \bibfield  {author} {\bibinfo {author} {\bibfnamefont {R.}~\bibnamefont
  {Cheaito}}, \bibinfo {author} {\bibfnamefont {J.~C.}\ \bibnamefont {Duda}},
  \bibinfo {author} {\bibfnamefont {T.~E.}\ \bibnamefont {Beechem}}, \bibinfo
  {author} {\bibfnamefont {K.}~\bibnamefont {Hattar}}, \bibinfo {author}
  {\bibfnamefont {J.~F.}\ \bibnamefont {Ihlefeld}}, \bibinfo {author}
  {\bibfnamefont {D.~L.}\ \bibnamefont {Medlin}}, \bibinfo {author}
  {\bibfnamefont {M.~A.}\ \bibnamefont {Rodriguez}}, \bibinfo {author}
  {\bibfnamefont {M.~J.}\ \bibnamefont {Campion}}, \bibinfo {author}
  {\bibfnamefont {E.~S.}\ \bibnamefont {Piekos}},\ and\ \bibinfo {author}
  {\bibfnamefont {P.~E.}\ \bibnamefont {Hopkins}},\ }\href
  {https://journals.aps.org/prl/abstract/10.1103/PhysRevLett.109.195901}
  {\bibfield  {journal} {\bibinfo  {journal} {Phys. Rev. Lett.}\ }\textbf
  {\bibinfo {volume} {109}},\ \bibinfo {pages} {195901} (\bibinfo {year}
  {2012})}\BibitemShut {NoStop}%
\bibitem [{\citenamefont {Braun}\ \emph {et~al.}(2016)\citenamefont {Braun},
  \citenamefont {Baker}, \citenamefont {Giri}, \citenamefont {Elahi},
  \citenamefont {Artyushkova}, \citenamefont {Beechem}, \citenamefont {Norris},
  \citenamefont {Leseman}, \citenamefont {Gaskins},\ and\ \citenamefont
  {Hopkins}}]{braun2016size}%
  \BibitemOpen
  \bibfield  {author} {\bibinfo {author} {\bibfnamefont {J.~L.}\ \bibnamefont
  {Braun}}, \bibinfo {author} {\bibfnamefont {C.~H.}\ \bibnamefont {Baker}},
  \bibinfo {author} {\bibfnamefont {A.}~\bibnamefont {Giri}}, \bibinfo {author}
  {\bibfnamefont {M.}~\bibnamefont {Elahi}}, \bibinfo {author} {\bibfnamefont
  {K.}~\bibnamefont {Artyushkova}}, \bibinfo {author} {\bibfnamefont {T.~E.}\
  \bibnamefont {Beechem}}, \bibinfo {author} {\bibfnamefont {P.~M.}\
  \bibnamefont {Norris}}, \bibinfo {author} {\bibfnamefont {Z.~C.}\
  \bibnamefont {Leseman}}, \bibinfo {author} {\bibfnamefont {J.~T.}\
  \bibnamefont {Gaskins}},\ and\ \bibinfo {author} {\bibfnamefont {P.~E.}\
  \bibnamefont {Hopkins}},\ }\href
  {https://journals.aps.org/prb/abstract/10.1103/PhysRevB.93.140201} {\bibfield
   {journal} {\bibinfo  {journal} {Phys. Rev. B}\ }\textbf {\bibinfo {volume}
  {93}},\ \bibinfo {pages} {140201} (\bibinfo {year} {2016})}\BibitemShut
  {NoStop}%
\bibitem [{\citenamefont {Li}\ \emph {et~al.}(2012)\citenamefont {Li},
  \citenamefont {Lindsay}, \citenamefont {Broido}, \citenamefont {Stewart},\
  and\ \citenamefont {Mingo}}]{li2012thermal}%
  \BibitemOpen
  \bibfield  {author} {\bibinfo {author} {\bibfnamefont {W.}~\bibnamefont
  {Li}}, \bibinfo {author} {\bibfnamefont {L.}~\bibnamefont {Lindsay}},
  \bibinfo {author} {\bibfnamefont {D.~A.}\ \bibnamefont {Broido}}, \bibinfo
  {author} {\bibfnamefont {D.~A.}\ \bibnamefont {Stewart}},\ and\ \bibinfo
  {author} {\bibfnamefont {N.}~\bibnamefont {Mingo}},\ }\href
  {https://journals.aps.org/prb/abstract/10.1103/PhysRevB.86.174307} {\bibfield
   {journal} {\bibinfo  {journal} {Phys. Rev. B}\ }\textbf {\bibinfo {volume}
  {86}},\ \bibinfo {pages} {174307} (\bibinfo {year} {2012})}\BibitemShut
  {NoStop}%
\bibitem [{\citenamefont {Feng}\ and\ \citenamefont
  {Ruan}(2016)}]{feng2016ultra}%
  \BibitemOpen
  \bibfield  {author} {\bibinfo {author} {\bibfnamefont {T.}~\bibnamefont
  {Feng}}\ and\ \bibinfo {author} {\bibfnamefont {X.}~\bibnamefont {Ruan}},\
  }\href
  {https://www.sciencedirect.com/science/article/pii/S0008622316300707?casa_token=JCuZakEizwgAAAAA:RS3LkXib6k3KTmkLfyaY-zztzNrzNDl0LiRVoj9Kq_4Fm3Gngvv9gPPLUr9gOzN2Td7aZXI}
  {\bibfield  {journal} {\bibinfo  {journal} {Carbon}\ }\textbf {\bibinfo
  {volume} {101}},\ \bibinfo {pages} {107} (\bibinfo {year}
  {2016})}\BibitemShut {NoStop}%
\bibitem [{\citenamefont {Perez-Taborda}\ \emph {et~al.}(2016)\citenamefont
  {Perez-Taborda}, \citenamefont {Rojo}, \citenamefont {Maiz}, \citenamefont
  {Neophytou},\ and\ \citenamefont {Martin-Gonzalez}}]{perez2016ultra}%
  \BibitemOpen
  \bibfield  {author} {\bibinfo {author} {\bibfnamefont {J.~A.}\ \bibnamefont
  {Perez-Taborda}}, \bibinfo {author} {\bibfnamefont {M.~M.}\ \bibnamefont
  {Rojo}}, \bibinfo {author} {\bibfnamefont {J.}~\bibnamefont {Maiz}}, \bibinfo
  {author} {\bibfnamefont {N.}~\bibnamefont {Neophytou}},\ and\ \bibinfo
  {author} {\bibfnamefont {M.}~\bibnamefont {Martin-Gonzalez}},\ }\href
  {https://www.nature.com/articles/srep32778} {\bibfield  {journal} {\bibinfo
  {journal} {Sci. Rep.}\ }\textbf {\bibinfo {volume} {6}},\ \bibinfo {pages}
  {1} (\bibinfo {year} {2016})}\BibitemShut {NoStop}%
\bibitem [{\citenamefont {Miura}\ \emph {et~al.}(2015)\citenamefont {Miura},
  \citenamefont {Zhou}, \citenamefont {Nozaki},\ and\ \citenamefont
  {Shiomi}}]{miura2015crystalline}%
  \BibitemOpen
  \bibfield  {author} {\bibinfo {author} {\bibfnamefont {A.}~\bibnamefont
  {Miura}}, \bibinfo {author} {\bibfnamefont {S.}~\bibnamefont {Zhou}},
  \bibinfo {author} {\bibfnamefont {T.}~\bibnamefont {Nozaki}},\ and\ \bibinfo
  {author} {\bibfnamefont {J.}~\bibnamefont {Shiomi}},\ }\href
  {https://pubs.acs.org/doi/abs/10.1021/acsami.5b02537?casa_token=HZkkLbPSip0AAAAA:XGd6OPryRkhA3diwNkzJnXx0mtNSuxZXKsLKmadeY93_2kcg2KXOjEgiLUCX5n3RqkdzMdELS8iz}
  {\bibfield  {journal} {\bibinfo  {journal} {ACS Appl. Mater. Interfaces}\
  }\textbf {\bibinfo {volume} {7}},\ \bibinfo {pages} {13484} (\bibinfo {year}
  {2015})}\BibitemShut {NoStop}%
\bibitem [{\citenamefont {Liao}\ and\ \citenamefont
  {Chen}(2015)}]{liao2015nanocomposites}%
  \BibitemOpen
  \bibfield  {author} {\bibinfo {author} {\bibfnamefont {B.}~\bibnamefont
  {Liao}}\ and\ \bibinfo {author} {\bibfnamefont {G.}~\bibnamefont {Chen}},\
  }\href {https://link.springer.com/article/10.1557/mrs.2015.197} {\bibfield
  {journal} {\bibinfo  {journal} {MRS Bull.}\ }\textbf {\bibinfo {volume}
  {40}},\ \bibinfo {pages} {746} (\bibinfo {year} {2015})}\BibitemShut
  {NoStop}%
\bibitem [{\citenamefont {Hosseini}\ \emph {et~al.}(2021)\citenamefont
  {Hosseini}, \citenamefont {Romano},\ and\ \citenamefont
  {Greaney}}]{doi:10.1021/acsaem.0c02640}%
  \BibitemOpen
  \bibfield  {author} {\bibinfo {author} {\bibfnamefont {S.~A.}\ \bibnamefont
  {Hosseini}}, \bibinfo {author} {\bibfnamefont {G.}~\bibnamefont {Romano}},\
  and\ \bibinfo {author} {\bibfnamefont {P.~A.}\ \bibnamefont {Greaney}},\
  }\href {https://doi.org/10.1021/acsaem.0c02640} {\bibfield  {journal}
  {\bibinfo  {journal} {ACS Appl. Energy Mater.}\ }\textbf {\bibinfo {volume}
  {4}},\ \bibinfo {pages} {1915} (\bibinfo {year} {2021})}\BibitemShut
  {NoStop}%
\bibitem [{\citenamefont {Shi}\ \emph {et~al.}(2018)\citenamefont {Shi},
  \citenamefont {Wu}, \citenamefont {Liu}, \citenamefont {Moshwan},
  \citenamefont {Wang}, \citenamefont {Chen},\ and\ \citenamefont
  {Zou}}]{shi2018polycrystalline}%
  \BibitemOpen
  \bibfield  {author} {\bibinfo {author} {\bibfnamefont {X.}~\bibnamefont
  {Shi}}, \bibinfo {author} {\bibfnamefont {A.}~\bibnamefont {Wu}}, \bibinfo
  {author} {\bibfnamefont {W.}~\bibnamefont {Liu}}, \bibinfo {author}
  {\bibfnamefont {R.}~\bibnamefont {Moshwan}}, \bibinfo {author} {\bibfnamefont
  {Y.}~\bibnamefont {Wang}}, \bibinfo {author} {\bibfnamefont {Z.-G.}\
  \bibnamefont {Chen}},\ and\ \bibinfo {author} {\bibfnamefont
  {J.}~\bibnamefont {Zou}},\ }\href
  {https://pubs.acs.org/doi/abs/10.1021/acsnano.8b06387?casa_token=9TZMkZdUo2EAAAAA:qhqeNpSsp6S1aayEF6Y5BaIa_yZM4SQav_2RJKe9mlgfwqxqdhXLBX2hcSuWQu6vnnv3p_6g64PM}
  {\bibfield  {journal} {\bibinfo  {journal} {ACS Nano}\ }\textbf {\bibinfo
  {volume} {12}},\ \bibinfo {pages} {11417} (\bibinfo {year}
  {2018})}\BibitemShut {NoStop}%
\bibitem [{\citenamefont {de~Sousa~Oliveira}\ and\ \citenamefont
  {Neophytou}(2019)}]{PhysRevB.100.035409}%
  \BibitemOpen
  \bibfield  {author} {\bibinfo {author} {\bibfnamefont {L.}~\bibnamefont
  {de~Sousa~Oliveira}}\ and\ \bibinfo {author} {\bibfnamefont {N.}~\bibnamefont
  {Neophytou}},\ }\href {https://doi.org/10.1103/PhysRevB.100.035409}
  {\bibfield  {journal} {\bibinfo  {journal} {Phys. Rev. B}\ }\textbf {\bibinfo
  {volume} {100}},\ \bibinfo {pages} {035409} (\bibinfo {year}
  {2019})}\BibitemShut {NoStop}%
\bibitem [{\citenamefont {de~Sousa~Oliveira}\ \emph {et~al.}(2020)\citenamefont
  {de~Sousa~Oliveira}, \citenamefont {Hosseini}, \citenamefont {Greaney},\ and\
  \citenamefont {Neophytou}}]{PhysRevB.102.205405}%
  \BibitemOpen
  \bibfield  {author} {\bibinfo {author} {\bibfnamefont {L.}~\bibnamefont
  {de~Sousa~Oliveira}}, \bibinfo {author} {\bibfnamefont {S.~A.}\ \bibnamefont
  {Hosseini}}, \bibinfo {author} {\bibfnamefont {A.}~\bibnamefont {Greaney}},\
  and\ \bibinfo {author} {\bibfnamefont {N.}~\bibnamefont {Neophytou}},\ }\href
  {https://doi.org/10.1103/PhysRevB.102.205405} {\bibfield  {journal} {\bibinfo
   {journal} {Phys. Rev. B}\ }\textbf {\bibinfo {volume} {102}},\ \bibinfo
  {pages} {205405} (\bibinfo {year} {2020})}\BibitemShut {NoStop}%
\bibitem [{\citenamefont {Lim}\ \emph {et~al.}(2016)\citenamefont {Lim},
  \citenamefont {Wang}, \citenamefont {Tang}, \citenamefont {Andrews},
  \citenamefont {So}, \citenamefont {Lee}, \citenamefont {Lee}, \citenamefont
  {Russell},\ and\ \citenamefont {Yang}}]{lim2016simultaneous}%
  \BibitemOpen
  \bibfield  {author} {\bibinfo {author} {\bibfnamefont {J.}~\bibnamefont
  {Lim}}, \bibinfo {author} {\bibfnamefont {H.-T.}\ \bibnamefont {Wang}},
  \bibinfo {author} {\bibfnamefont {J.}~\bibnamefont {Tang}}, \bibinfo {author}
  {\bibfnamefont {S.~C.}\ \bibnamefont {Andrews}}, \bibinfo {author}
  {\bibfnamefont {H.}~\bibnamefont {So}}, \bibinfo {author} {\bibfnamefont
  {J.}~\bibnamefont {Lee}}, \bibinfo {author} {\bibfnamefont {D.~H.}\
  \bibnamefont {Lee}}, \bibinfo {author} {\bibfnamefont {T.~P.}\ \bibnamefont
  {Russell}},\ and\ \bibinfo {author} {\bibfnamefont {P.}~\bibnamefont
  {Yang}},\ }\href {https://doi.org/10.1021/acsnano.5b05385} {\bibfield
  {journal} {\bibinfo  {journal} {ACS Nano}\ }\textbf {\bibinfo {volume}
  {10}},\ \bibinfo {pages} {124} (\bibinfo {year} {2016})}\BibitemShut
  {NoStop}%
\bibitem [{\citenamefont {Gurunathan}\ \emph
  {et~al.}(2020{\natexlab{a}})\citenamefont {Gurunathan}, \citenamefont
  {Hanus}, \citenamefont {Dylla}, \citenamefont {Katre},\ and\ \citenamefont
  {Snyder}}]{gurunathan2020analytical}%
  \BibitemOpen
  \bibfield  {author} {\bibinfo {author} {\bibfnamefont {R.}~\bibnamefont
  {Gurunathan}}, \bibinfo {author} {\bibfnamefont {R.}~\bibnamefont {Hanus}},
  \bibinfo {author} {\bibfnamefont {M.}~\bibnamefont {Dylla}}, \bibinfo
  {author} {\bibfnamefont {A.}~\bibnamefont {Katre}},\ and\ \bibinfo {author}
  {\bibfnamefont {G.~J.}\ \bibnamefont {Snyder}},\ }\href
  {https://doi.org/10.1103/PhysRevApplied.13.034011} {\bibfield  {journal}
  {\bibinfo  {journal} {Phys. Rev. Applied}\ }\textbf {\bibinfo {volume}
  {13}},\ \bibinfo {pages} {034011} (\bibinfo {year}
  {2020}{\natexlab{a}})}\BibitemShut {NoStop}%
\bibitem [{\citenamefont {Al~Rahal Al~Orabi}\ \emph {et~al.}(2016)\citenamefont
  {Al~Rahal Al~Orabi}, \citenamefont {Mecholsky}, \citenamefont {Hwang},
  \citenamefont {Kim}, \citenamefont {Rhyee}, \citenamefont {Wee},\ and\
  \citenamefont {Fornari}}]{al2016band}%
  \BibitemOpen
  \bibfield  {author} {\bibinfo {author} {\bibfnamefont {R.}~\bibnamefont
  {Al~Rahal Al~Orabi}}, \bibinfo {author} {\bibfnamefont {N.~A.}\ \bibnamefont
  {Mecholsky}}, \bibinfo {author} {\bibfnamefont {J.}~\bibnamefont {Hwang}},
  \bibinfo {author} {\bibfnamefont {W.}~\bibnamefont {Kim}}, \bibinfo {author}
  {\bibfnamefont {J.-S.}\ \bibnamefont {Rhyee}}, \bibinfo {author}
  {\bibfnamefont {D.}~\bibnamefont {Wee}},\ and\ \bibinfo {author}
  {\bibfnamefont {M.}~\bibnamefont {Fornari}},\ }\href
  {https://doi.org/10.1021/acs.chemmater.5b04365} {\bibfield  {journal}
  {\bibinfo  {journal} {Chem. Mater.}\ }\textbf {\bibinfo {volume} {28}},\
  \bibinfo {pages} {376} (\bibinfo {year} {2016})}\BibitemShut {NoStop}%
\bibitem [{\citenamefont {Gurunathan}\ \emph
  {et~al.}(2020{\natexlab{b}})\citenamefont {Gurunathan}, \citenamefont
  {Hanus},\ and\ \citenamefont {Snyder}}]{gurunathan2020alloy}%
  \BibitemOpen
  \bibfield  {author} {\bibinfo {author} {\bibfnamefont {R.}~\bibnamefont
  {Gurunathan}}, \bibinfo {author} {\bibfnamefont {R.}~\bibnamefont {Hanus}},\
  and\ \bibinfo {author} {\bibfnamefont {G.~J.}\ \bibnamefont {Snyder}},\
  }\href {https://doi.org/10.1039/C9MH01990A} {\bibfield  {journal} {\bibinfo
  {journal} {Mater. Horiz.}\ }\textbf {\bibinfo {volume} {7}},\ \bibinfo
  {pages} {1452} (\bibinfo {year} {2020}{\natexlab{b}})}\BibitemShut {NoStop}%
\bibitem [{\citenamefont {Ju}\ \emph {et~al.}(2017)\citenamefont {Ju},
  \citenamefont {Kim}, \citenamefont {Park},\ and\ \citenamefont
  {Kim}}]{doi:10.1021/acs.chemmater.7b00423}%
  \BibitemOpen
  \bibfield  {author} {\bibinfo {author} {\bibfnamefont {H.}~\bibnamefont
  {Ju}}, \bibinfo {author} {\bibfnamefont {M.}~\bibnamefont {Kim}}, \bibinfo
  {author} {\bibfnamefont {D.}~\bibnamefont {Park}},\ and\ \bibinfo {author}
  {\bibfnamefont {J.}~\bibnamefont {Kim}},\ }\href
  {https://doi.org/10.1021/acs.chemmater.7b00423} {\bibfield  {journal}
  {\bibinfo  {journal} {Chem. Mater.}\ }\textbf {\bibinfo {volume} {29}},\
  \bibinfo {pages} {3228} (\bibinfo {year} {2017})}\BibitemShut {NoStop}%
\bibitem [{\citenamefont {Bera}\ \emph {et~al.}(2010)\citenamefont {Bera},
  \citenamefont {Mingo},\ and\ \citenamefont {Volz}}]{bera2010marked}%
  \BibitemOpen
  \bibfield  {author} {\bibinfo {author} {\bibfnamefont {C.}~\bibnamefont
  {Bera}}, \bibinfo {author} {\bibfnamefont {N.}~\bibnamefont {Mingo}},\ and\
  \bibinfo {author} {\bibfnamefont {S.}~\bibnamefont {Volz}},\ }\href
  {https://doi.org/10.1103/PhysRevLett.104.115502} {\bibfield  {journal}
  {\bibinfo  {journal} {Phys. Rev. Lett.}\ }\textbf {\bibinfo {volume} {104}},\
  \bibinfo {pages} {115502} (\bibinfo {year} {2010})}\BibitemShut {NoStop}%
\bibitem [{\citenamefont {Yang}\ and\ \citenamefont
  {Minnich}(2017)}]{yang2017thermal}%
  \BibitemOpen
  \bibfield  {author} {\bibinfo {author} {\bibfnamefont {L.}~\bibnamefont
  {Yang}}\ and\ \bibinfo {author} {\bibfnamefont {A.~J.}\ \bibnamefont
  {Minnich}},\ }\href@noop {} {\bibfield  {journal} {\bibinfo  {journal}
  {Scientific reports}\ }\textbf {\bibinfo {volume} {7}},\ \bibinfo {pages} {1}
  (\bibinfo {year} {2017})}\BibitemShut {NoStop}%
\bibitem [{\citenamefont {Romano}(2021)}]{romano2021}%
  \BibitemOpen
  \bibfield  {author} {\bibinfo {author} {\bibfnamefont {G.}~\bibnamefont
  {Romano}},\ }\href {https://arxiv.org/pdf/2105.08181.pdf} {\bibfield
  {journal} {\bibinfo  {journal} {Arxiv}\ } (\bibinfo {year}
  {2021})}\BibitemShut {NoStop}%
\bibitem [{\citenamefont {Harter}\ \emph {et~al.}(2019)\citenamefont {Harter},
  \citenamefont {Hosseini}, \citenamefont {Palmer},\ and\ \citenamefont
  {Greaney}}]{harter2019prediction}%
  \BibitemOpen
  \bibfield  {author} {\bibinfo {author} {\bibfnamefont {J.~R.}\ \bibnamefont
  {Harter}}, \bibinfo {author} {\bibfnamefont {S.~A.}\ \bibnamefont
  {Hosseini}}, \bibinfo {author} {\bibfnamefont {T.~S.}\ \bibnamefont
  {Palmer}},\ and\ \bibinfo {author} {\bibfnamefont {P.~A.}\ \bibnamefont
  {Greaney}},\ }\href
  {https://www.sciencedirect.com/science/article/pii/S0017931019325529?casa_token=t0IWZQUBvwsAAAAA:SZ-vwpyupjPoAeuaMXTwgJ0aTPNrtIZs26C2g4Y9mLtZrN22xTvJ3zK0hHVGycn0Auw5bok}
  {\bibfield  {journal} {\bibinfo  {journal} {Int. J. Heat Mass Transf.}\
  }\textbf {\bibinfo {volume} {144}},\ \bibinfo {pages} {118595} (\bibinfo
  {year} {2019})}\BibitemShut {NoStop}%
\bibitem [{\citenamefont {Harter}\ \emph {et~al.}(2020)\citenamefont {Harter},
  \citenamefont {Palmer},\ and\ \citenamefont
  {Greaney}}]{harter2020predicting}%
  \BibitemOpen
  \bibfield  {author} {\bibinfo {author} {\bibfnamefont {J.~R.}\ \bibnamefont
  {Harter}}, \bibinfo {author} {\bibfnamefont {T.~S.}\ \bibnamefont {Palmer}},\
  and\ \bibinfo {author} {\bibfnamefont {P.~A.}\ \bibnamefont {Greaney}},\
  }\href
  {https://www.researchgate.net/profile/Jackson-Harter/publication/344468065_Predicting_mesoscale_spectral_thermal_conductivity_using_advanced_deterministic_phonon_transport_techniques/links/5fa1894192851c14bcff797a/Predicting-mesoscale-spectral-thermal-conductivity-using-advanced-deterministic-phonon-transport-techniques.pdf}
  {\bibfield  {journal} {\bibinfo  {journal} {Adv. Heat Transf.}\ }\textbf
  {\bibinfo {volume} {52}} (\bibinfo {year} {2020})}\BibitemShut {NoStop}%
\bibitem [{\citenamefont {Ziman}(1965)}]{ziman1972principles}%
  \BibitemOpen
  \bibfield  {author} {\bibinfo {author} {\bibfnamefont {J.~M.}\ \bibnamefont
  {Ziman}},\ }\href {https://doi.org/10.1119/1.1971507} {\bibfield  {journal}
  {\bibinfo  {journal} {Am. J. Phys.}\ }\textbf {\bibinfo {volume} {33}},\
  \bibinfo {pages} {349} (\bibinfo {year} {1965})}\BibitemShut {NoStop}%
\bibitem [{\citenamefont {Carrete}\ \emph {et~al.}(2017)\citenamefont
  {Carrete}, \citenamefont {Vermeersch}, \citenamefont {Katre}, \citenamefont
  {van Roekeghem}, \citenamefont {Wang}, \citenamefont {Madsen},\ and\
  \citenamefont {Mingo}}]{carrete2017almabte}%
  \BibitemOpen
  \bibfield  {author} {\bibinfo {author} {\bibfnamefont {J.}~\bibnamefont
  {Carrete}}, \bibinfo {author} {\bibfnamefont {B.}~\bibnamefont {Vermeersch}},
  \bibinfo {author} {\bibfnamefont {A.}~\bibnamefont {Katre}}, \bibinfo
  {author} {\bibfnamefont {A.}~\bibnamefont {van Roekeghem}}, \bibinfo {author}
  {\bibfnamefont {T.}~\bibnamefont {Wang}}, \bibinfo {author} {\bibfnamefont
  {G.~K.}\ \bibnamefont {Madsen}},\ and\ \bibinfo {author} {\bibfnamefont
  {N.}~\bibnamefont {Mingo}},\ }\href
  {https://www.sciencedirect.com/science/article/pii/S0010465517302059?casa_token=gwNF0U_SqqUAAAAA:xrP5ntN7mNYefTHP6cCIjtnYh7YZOxdOzA_qQ18yUic40aqJIusgHu2yUI_hWGAIPniKV2Y}
  {\bibfield  {journal} {\bibinfo  {journal} {Comput. Phys. Commun.}\ }\textbf
  {\bibinfo {volume} {220}},\ \bibinfo {pages} {351} (\bibinfo {year}
  {2017})}\BibitemShut {NoStop}%
\bibitem [{\citenamefont {Tamura}(1983)}]{tamura1983isotope}%
  \BibitemOpen
  \bibfield  {author} {\bibinfo {author} {\bibfnamefont {S.-i.}\ \bibnamefont
  {Tamura}},\ }\href {https://doi.org/10.1103/PhysRevB.27.858} {\bibfield
  {journal} {\bibinfo  {journal} {Phys. Rev. B}\ }\textbf {\bibinfo {volume}
  {27}},\ \bibinfo {pages} {858} (\bibinfo {year} {1983})}\BibitemShut
  {NoStop}%
\bibitem [{\citenamefont {Arrigoni}\ \emph {et~al.}(2018)\citenamefont
  {Arrigoni}, \citenamefont {Carrete}, \citenamefont {Mingo},\ and\
  \citenamefont {Madsen}}]{arrigoni2018first}%
  \BibitemOpen
  \bibfield  {author} {\bibinfo {author} {\bibfnamefont {M.}~\bibnamefont
  {Arrigoni}}, \bibinfo {author} {\bibfnamefont {J.}~\bibnamefont {Carrete}},
  \bibinfo {author} {\bibfnamefont {N.}~\bibnamefont {Mingo}},\ and\ \bibinfo
  {author} {\bibfnamefont {G.~K.~H.}\ \bibnamefont {Madsen}},\ }\href
  {https://doi.org/10.1103/PhysRevB.98.115205} {\bibfield  {journal} {\bibinfo
  {journal} {Phys. Rev. B}\ }\textbf {\bibinfo {volume} {98}},\ \bibinfo
  {pages} {115205} (\bibinfo {year} {2018})}\BibitemShut {NoStop}%
\bibitem [{\citenamefont {Kim}(2015)}]{kim2015strategies}%
  \BibitemOpen
  \bibfield  {author} {\bibinfo {author} {\bibfnamefont {W.}~\bibnamefont
  {Kim}},\ }\href {https://doi.org/10.1039/C5TC01670C} {\bibfield  {journal}
  {\bibinfo  {journal} {J. Mater. Chem. C}\ }\textbf {\bibinfo {volume} {3}},\
  \bibinfo {pages} {10336} (\bibinfo {year} {2015})}\BibitemShut {NoStop}%
\bibitem [{\citenamefont {Li}\ \emph {et~al.}(2014)\citenamefont {Li},
  \citenamefont {Carrete}, \citenamefont {Katcho},\ and\ \citenamefont
  {Mingo}}]{li2014shengbte}%
  \BibitemOpen
  \bibfield  {author} {\bibinfo {author} {\bibfnamefont {W.}~\bibnamefont
  {Li}}, \bibinfo {author} {\bibfnamefont {J.}~\bibnamefont {Carrete}},
  \bibinfo {author} {\bibfnamefont {N.~A.}\ \bibnamefont {Katcho}},\ and\
  \bibinfo {author} {\bibfnamefont {N.}~\bibnamefont {Mingo}},\ }\href
  {https://www.sciencedirect.com/science/article/pii/S0010465514000484?casa_token=0tzPWEO4G0EAAAAA:6VxP9CinkEIrBXA7IGMOa9uiaDirVQaCCuso7onlS30ErCAh33gPe8lgEC6IK1rqKXVSMG0}
  {\bibfield  {journal} {\bibinfo  {journal} {Comput. Phys. Commun.}\ }\textbf
  {\bibinfo {volume} {185}},\ \bibinfo {pages} {1747} (\bibinfo {year}
  {2014})}\BibitemShut {NoStop}%
\bibitem [{\citenamefont {Xie}\ \emph {et~al.}(2013)\citenamefont {Xie},
  \citenamefont {Wang}, \citenamefont {Pei}, \citenamefont {Fu}, \citenamefont
  {Liu}, \citenamefont {Snyder}, \citenamefont {Zhao},\ and\ \citenamefont
  {Zhu}}]{xie2013beneficial}%
  \BibitemOpen
  \bibfield  {author} {\bibinfo {author} {\bibfnamefont {H.}~\bibnamefont
  {Xie}}, \bibinfo {author} {\bibfnamefont {H.}~\bibnamefont {Wang}}, \bibinfo
  {author} {\bibfnamefont {Y.}~\bibnamefont {Pei}}, \bibinfo {author}
  {\bibfnamefont {C.}~\bibnamefont {Fu}}, \bibinfo {author} {\bibfnamefont
  {X.}~\bibnamefont {Liu}}, \bibinfo {author} {\bibfnamefont {G.~J.}\
  \bibnamefont {Snyder}}, \bibinfo {author} {\bibfnamefont {X.}~\bibnamefont
  {Zhao}},\ and\ \bibinfo {author} {\bibfnamefont {T.}~\bibnamefont {Zhu}},\
  }\href {https://doi.org/https://doi.org/10.1002/adfm.201300663} {\bibfield
  {journal} {\bibinfo  {journal} {Adv. Funct. Mater.}\ }\textbf {\bibinfo
  {volume} {23}},\ \bibinfo {pages} {5123} (\bibinfo {year}
  {2013})}\BibitemShut {NoStop}%
\bibitem [{\citenamefont {Toberer}\ \emph {et~al.}(2011)\citenamefont
  {Toberer}, \citenamefont {Zevalkink},\ and\ \citenamefont
  {Snyder}}]{toberer2011phonon}%
  \BibitemOpen
  \bibfield  {author} {\bibinfo {author} {\bibfnamefont {E.~S.}\ \bibnamefont
  {Toberer}}, \bibinfo {author} {\bibfnamefont {A.}~\bibnamefont {Zevalkink}},\
  and\ \bibinfo {author} {\bibfnamefont {G.~J.}\ \bibnamefont {Snyder}},\
  }\href {https://doi.org/10.1039/C1JM11754H} {\bibfield  {journal} {\bibinfo
  {journal} {J. Mater. Chem.}\ }\textbf {\bibinfo {volume} {21}},\ \bibinfo
  {pages} {15843} (\bibinfo {year} {2011})}\BibitemShut {NoStop}%
\bibitem [{\citenamefont {Romano}\ and\ \citenamefont
  {Kolpak}(2019)}]{romano2019diffusive}%
  \BibitemOpen
  \bibfield  {author} {\bibinfo {author} {\bibfnamefont {G.}~\bibnamefont
  {Romano}}\ and\ \bibinfo {author} {\bibfnamefont {A.~M.}\ \bibnamefont
  {Kolpak}},\ }\href
  {https://asmedigitalcollection.asme.org/heattransfer/article/141/1/012401/365762/Diffusive-Phonons-in-Nongray-Nanostructures?casa_token=B3aThsn5JgkAAAAA:doIXTlSelk6sClt2zOj2ON9_Eou-rSZRER0GqDx-vgZpzZZri9M2nYt2ovj-6wkXc4UN}
  {\bibfield  {journal} {\bibinfo  {journal} {J Heat Trans.}\ }\textbf
  {\bibinfo {volume} {141}} (\bibinfo {year} {2019})}\BibitemShut {NoStop}%
\bibitem [{\citenamefont {Hosseini}(2021)}]{Hosseini2021}%
  \BibitemOpen
  \bibfield  {author} {\bibinfo {author} {\bibfnamefont {S.~A.}\ \bibnamefont
  {Hosseini}},\ }\href@noop {} {\bibinfo {title}
  {Universal-effective-medium-theory}},\ \bibinfo {howpublished}
  {https://github.com/ariahosseini/Universal-Effective-Medium-Theory} (\bibinfo
  {year} {2021})\BibitemShut {NoStop}%
\end{thebibliography}%

\end{document}